# Broadband Unidirectional Visible Imaging Using Wafer-Scale Nano-Fabrication of Multi-Layer Diffractive Optical Processors

Che-Yung Shen[1,2,3], Paolo Batoni[4], Xilin Yang[1,2,3], Jingxi Li[1,2,3], Kun Liao[1], Jared Stack[4], Jeff Gardner[4], Kevin Welch[4] and Aydogan Ozcan[1,2,3*]

[1]Electrical and Computer Engineering Department, University of California, Los Angeles, CA, 90095, USA

[2]Bioengineering Department, University of California, Los Angeles, CA, 90095, USA

[3]California NanoSystems Institute (CNSI), University of California, Los Angeles, CA, 90095, USA

[4]Micro Optics, Optical Systems Division, Broadcom Inc., Charlotte, NC, 28262, USA

[*]Correspondence to: ozcan@ucla.edu

## Abstract

We present a broadband and polarization-insensitive unidirectional imager that operates at the visible part of the spectrum, where image formation occurs in one direction while in the opposite direction, it is blocked. This approach is enabled by deep learning-driven diffractive optical design with wafer-scale nano-fabrication using high-purity fused silica to ensure optical transparency and thermal stability. Our design achieves unidirectional imaging across three visible wavelengths (covering red, green and blue parts of the spectrum), and we experimentally validated this broadband unidirectional imager by creating high-fidelity images in the forward direction and generating weak, distorted output patterns in the backward direction, in alignment with our numerical simulations. This work demonstrates the wafer-scale production of diffractive optical processors, featuring 16 levels of nanoscale phase features distributed across two axially aligned diffractive layers for visible unidirectional imaging. This approach facilitates mass-scale production of ~0.5 billion nanoscale phase features per wafer, supporting high-throughput manufacturing of hundreds to thousands of multi-layer diffractive processors suitable for large apertures and parallel processing of multiple tasks. Our design can seamlessly integrate into conventional optical systems, broadening its applicability in fields such as security, defense, and telecommunication. Beyond broadband unidirectional imaging in the visible spectrum, this study establishes a pathway for artificial-intelligence-enabled diffractive optics with versatile applications, signaling a new era in optical device functionality with industrial-level massively scalable fabrication.

## INTRODUCTION

Deep learning has been transforming optical engineering by enabling novel approaches to the inverse design of optical systems[1–15]. For example, deep learning-driven inverse design of diffractive optical elements (DOEs) has led to the development of spatially-engineered diffractive layers, forming various architectures of diffractive optical processors[16–20] where multi-layer diffractive structures collectively execute different target functions. These diffractive processors, composed of cascaded layers with wavelength-scale features, allow precise modulation of optical fields to achieve a wide range of advanced tasks, including quantitative phase imaging[21–24], all-optical phase conjugation[25], image denoising[26], spectral filtering[27–29], and class-specific imaging[30,31]. Metasurfaces[32–34], as another example, utilize deeply subwavelength features to achieve customized optical responses, allowing precise control over various light properties, including phase, polarization, dispersion, and orbital angular momentum. These innovations in spatially engineered surfaces represent a significant advancement in optical information processing, enabling various applications, including beam steering[35], holography[36,37], space-efficient optical computing[38] and smart imaging[39,40].

Despite the promising potential and emerging uses of diffractive optical processors and metamaterials, most of these demonstrations remain constrained to 2D implementations and longer wavelengths due to the fabrication challenges of nanoscale features in 3D diffractive architectures. Nano-fabrication techniques such as two-photon polymerization methods[9,31,41,42] and electron beam lithography (EBL) processes[14,32,40,43,44] have enabled the fabrication of micro- or nanoscale multi-layer diffractive designs for the near-infrared and visible spectrum. However, these designs suffer from material absorption, restricted fabrication area, limited phase bit depth for each modulation element, and 3D alignment challenges, resulting in limited degrees of freedom for more complex applications in the visible spectrum. To our knowledge, no prior demonstration reports wafer-level scalable fabrication of multi-layer diffractive surfaces operating at visible wavelengths.

Here, we demonstrate multi-layer diffractive optical processors for broadband unidirectional imaging in the visible spectrum using industrial-grade on-wafer lithography[45,46]; **Figure 1**. Notably, our design features the scalable fabrication of two-layer diffractive optical processors, specifically designed for the visible spectrum, achieving a 16-level phase depth per diffractive nanoscale feature. This all-optical diffractive processor enables visible image formation in only one direction—transmitting images from input field of view (FOV) A to output FOV B—while blocking and distorting image formation in the reverse direction (B → A). This work represents the first demonstration of broadband unidirectional imaging in the visible spectrum, achieved with nanoscale, polarization-insensitive and transparent diffractive features that were optimized using deep learning.

We experimentally validated the unidirectional imaging capability of our two-layer diffractive processor across three wavelengths: 467.5 nm, 525 nm and 627.5 nm, corresponding to blue, green and red colors. The unidirectional imager successfully created the forward-direction images while the backward-direction information was blocked and distorted for all three illumination wavelengths, well matching the numerical simulations. Despite being trained with three narrow spectral bands, the diffractive imager maintained successful unidirectional image transmission over a continuum of wavelengths, demonstrating robustness as a broadband unidirectional imager.

Our 3D nano-fabrication method supports scalable, high-throughput manufacturing of hundreds of millions of phase features on the same wafer, making it suitable for large FOV operation over extended apertures and parallel multi-task processing. Coupled with the use of high-purity fused silica (HPFS)—highly valuable for its ultra-low energy loss and exceptional thermal stability—this advancement enables complex diffractive processing, making our multi-layer designs adaptable for a wide range of optical applications. Given that our fabrication methods overlap with the lithography processes used in semiconductor manufacturing, our diffractive optical processor designs could be monolithically integrated with other electronic or optoelectronic devices. The potential applications for diffractive unidirectional imaging using structured materials are extensive, spanning fields such as security, defense, telecommunications, and privacy protection. This study opens up new avenues for the applications of diffractive optical processors, paving the way for advanced, massively scalable solutions in intelligent imaging and sensing with visible light.

**BROADBAND UNIDIRECTIONAL IMAGER DESIGN**

**Figure 1a** depicts a schematic of our broadband unidirectional imaging framework, illuminated by spatially coherent light across different wavelengths. This system comprises input/output FOVs and a diffractive imaging unit with two successive modulation layers that are structured transmissive surfaces. Each diffractive layer consists of 512×512 trainable diffractive features, with each feature having a lateral size of 714 nm and a tunable thickness, providing a phase modulation range covering 0-2π for all the desired illumination wavelengths. The two transmissive layers are made of and connected through HPFS, with both the input plane to the first diffractive layer and the second diffractive layer to the output plane interfaced via light diffraction in air; this configuration results in an axially compact system spanning ~2 mm. In addition to thermal stability, the choice of HPFS offers several other key benefits. It is mechanically strong, highly resistant to abrasion, chemically inert, and resilient to both strong acids and bases, making it suitable for harsh environments. Furthermore, HPFS is available in standard SEMI wafer form with double-sided optical polish, facilitating compatibility with wafer-scale manufacturing processes. Unlike conventional glass compositions, HPFS is a pure, amorphous form of $SiO_2$, allowing for precise dry etching with standard chemistries without dependence on crystal orientation (see the Methods section for details).

The broadband unidirectional imager processes the complex fields of the multispectral input object $\{i_w\}$ to produce output complex fields $\{o_w\}$ at each wavelength of interest ($\lambda_w$). The resulting output intensity profiles are captured in a single snapshot by a color Complementary Metal-Oxide-Semiconductor (CMOS) image sensor, providing intensity measurements $\boldsymbol{O}_w$. In the forward direction, the unidirectional imager faithfully reproduces the corresponding image at each wavelength within the output FOV, i.e., $\boldsymbol{O}_w \approx \alpha_w \cdot \boldsymbol{I}_w$, where $\boldsymbol{I}_w$ is the ground truth intensity pattern of the input object and $\alpha_w$ is a wavelength-dependent scalar constant. In the backward direction, when the input and output FOVs are reversed, the unidirectional imager blocks the image information, yielding distorted, reduced-energy patterns at the output FOV across all the desired illumination wavelengths.

Our diffractive unidirectional imaging models were optimized using error backpropagation and stochastic gradient descent-based optimization[47], aiming to minimize a custom loss function (*L*) based on the normalized mean-squared error (NMSE), Pearson Correlation Coefficient (PCC), and diffraction efficiency between the projected intensity images and their corresponding ground truth images across all the wavelengths – calculated for both the forward and backward directions (see the Methods section for details). To achieve broadband unidirectional imaging capability in the visible spectrum, the system was trained using wavelengths randomly sampled within {627.5 ± 10 nm, 525 ± 18 nm, 467.5 ± 7.5 nm} during each training iteration as detailed in the Methods section. Deep learning-based training used the MNIST image dataset, and the resulting optimized diffractive layers, with 16 levels of phase for each diffractive modulation element, are shown on the left side of **Fig. 1b**.

To evaluate the broadband imaging capability of this unidirectional imager design, we conducted a numerical analysis of its spectral response across the visible spectrum, spanning [450 – 650] nm with 200 uniformly sampled test wavelengths. This analysis compared the unidirectional imaging performances of two diffractive designs: (i) the two-layer unidirectional imager design (shown in **Fig. 1b**) and (ii) a three-layer imager design that incorporates an additional diffractive layer for enhanced spectral response (shown in **Supplementary Fig. 1**). The latter design, except for the number of diffractive layers, retained all the other structural parameters identical to the former two-layer configuration and utilized the same training image dataset. To assess these broadband unidirectional imagers' internal and external generalization capabilities, we numerically tested each design using 10,000 input images from the MNIST dataset and 10,000 input images from the Fashion MNIST dataset, both of which were never used during the training stage. **Figure 2** shows examples of the output images resulting from both of these unidirectional imager designs. As illustrated in **Fig. 2a**, the two-layer diffractive unidirectional imager successfully reproduces forward images while significantly distorting and blocking the backward image formation, as desired. This asymmetrical image transmission remained successful across different input datasets, including the MNIST and Fashion MNIST datasets, demonstrating the unidirectional imager's generalization capability to different types of input objects. The three-layer unidirectional imager design, shown in **Fig. 2b**, achieved further improved performance, not only enhancing the quality of the forward images but also further suppressing the undesired image formation in the backward direction.

Quantitative metrics reported in **Fig. 3** further validated the performances of both of these diffractive unidirectional imager designs. As shown in **Fig. 3a**, the two-layer imager consistently maintained effective unidirectional imaging performance across different wavelengths, achieving forward PCC values of > 0.86 and backward PCC values of < 0.58 throughout the tested illumination spectrum, 450 – 650 nm. This shallower design with 2 diffractive layers also demonstrated asymmetrical energy transmission, with forward diffraction efficiency values of > 28% and backward diffraction efficiency values of < 13% across all the tested wavelengths, as illustrated in **Fig. 3b**. The deeper diffractive unidirectional imager design with 3 layers provided significant enhancements in these metrics. As shown in **Fig. 3a**, the forward PCC values of the deeper design with 3 optimized diffractive layers were improved to > 0.89 across the entire spectrum, while the backward PCC values dropped to < 0.33, demonstrating a substantial improvement in suppressing undesired image formation in the backward direction. Furthermore, **Fig. 3b** revealed that the forward diffraction efficiency increased to > 30%, while the backward efficiency dropped to < 10% across the entire test wavelength range. This increased performance asymmetry between the forward and backward directions highlights the superior capability of deeper diffractive processor designs with more degrees of freedom for enhanced unidirectional imaging.

**WAFER-SCALE FABRICATION OF BROADBAND UNIDIRECTIONAL VISIBLE IMAGERS**

**Figure 1b** illustrates the wafer-scale fabrication of diffractive optical processors specifically designed for broadband unidirectional imaging in the visible part of the spectrum. We also highlight the zoomed-in diffractive layers and the individual nanoscale diffractive features in **Fig. 1b**. Using wafer-scale manufacturing, we fabricated 918 multi-layer diffractive designs on a single wafer, exhibiting ~0.5 billion 16-phase-level diffractive features at the nanoscale.

To elaborate on the fabrication approach of our diffractive unidirectional visible imager, we first illustrate a generic manufacturing process, as shown in **Fig. 4**, which is used to create the binary architecture that characterizes DOEs. This structure is composed of discrete surface-relief micro and nanoscale "pixels"[48], and the manufacturing process behind it is closely related to the lithographic processes used to fabricate semiconductors at the wafer scale. First, a substrate wafer is coated with a thin, uniform layer of photoresist using a spin-coater. Depending on the desired minimum size, the photoresist may be patterned using either direct (1x) contact printing or a projection process such as a stepper or scanner (4x-5x). In this work, we utilized projection lithography to fabricate our diffractive design since the minimum feature size obtainable using projection lithography (~100 nm) is significantly smaller than that achievable using contact printing (~1 μm). After exposure using a suitable photomask, the resist is developed and fully cured. A permanent pattern on the substrate's surface is then made through etching, which removes the exposed areas of the wafer. As an irreversible step, etching is the most critical phase of the process.

For manufacturing distinct geometries and vertical sidewalls, dry etching[49], typically using gas-phase chemical reactions, is preferred over wet etching, which uses amorphous materials with liquid chemistry. As shown in **Fig. 4a**, the advantage of dry etching is that it can be made anisotropic (unidirectional), meaning that etching takes place at a high etch rate in one direction, typically in a vertical direction with respect to the surface of the substrate wafer. This dry etching method preserves the lateral size and shape of features. On the contrary, as shown in **Fig. 4b**, wet etching proceeds with the same etch rate in every direction (omnidirectional). This isotropy typically causes broadening of feature dimensions and rounding of edges, limiting the precision of fabricated structures. Given these characteristics, our fabrication process employed chlorine-based dry etching chemistries to achieve precisely controlled axial depths. This approach enabled us to produce vertical profiles without altering the diffractive elements' lateral dimensions, ensuring each feature's structural integrity and optical performance.

We achieved the fabrication of 16-phase-level diffractive structures on two opposing surfaces of our 6-inch HPFS wafer (see the Methods section). This type of architecture can be generalized by repeating a 2-level design concept. In this approach, the fabrication involves repeating a cycle of coating, exposing, developing and etching multiple times with a customized photomask for each level. **Fig. 4c-f** illustrates this process, showing how a blazed structure can be approximated by successively etching smaller steps with precise relative alignment. Each step size is controlled solely by the etching time, provided that the etching rate under specific operating conditions is known. Patterning of the photoresist is achieved using UV exposure and different photomasks through a projection process. By repeating the fabrication cycle four times with four different photomasks, 16-phase level diffractive surfaces were successfully fabricated on the wafer.

**Figure 1b-c** provides a detailed look at both the numerically optimized diffractive thickness profiles and the experimentally fabricated layers. **Figure 1b** shows the wafer layout, with close to a thousand identical diffractive optical processor designs arranged in a 2D array on the same 6-inch wafer. High-resolution measurements obtained via scanning electron microscopy (SEM) (**Fig. 1c**) and confocal microscopy (**Supplementary Fig. 2**) reveal nanoscale accuracy in both the axial depth precision and lateral resolution. To evaluate the quantitative precision of the fabrication, etch depth error was measured on test structures across the wafer according to Six Sigma and International Organization for Standardization (ISO) standards[50,51]. Our unidirectional imager fabrication achieved an etch depth error within 3-5% across all diffractive designs across different wafers, ensuring consistent performance.

Our fabrication approach for multi-layer diffractive optical processors surpasses other methods, such as two-photon polymerization[9,31,42,52], optical Fourier surfaces[43,44], resin stamping[53] and mask-less grayscale exposure[54,55], in several critical aspects. It supports a significantly larger lateral area, improved axial resolution, and a phase depth of 16 levels, all while enabling wafer-level multi-layer fabrication. As shown in **Extended Data Table 1**, state-of-the-art two-photon polymerization-based 3D printing demonstrated the fabrication of multi-layer diffractive processors with a lateral feature size of ~400 nm and a minimum axial step size of ~10 nm, but this was limited to a total lateral size of <100 μm, restricting its applications. Similarly, the fabrication of optical surfaces using EBL demonstrated single-layer diffractive surfaces with a minimum axial step size of ~20 nm, but it was also limited to a total lateral size of <100 μm. Another fabrication approach using resin stamping demonstrated single-layer diffractive surfaces with a lateral feature size of 4 μm and an axial step size of ~160 nm, providing 8 thickness levels. Mask-less grayscale exposure, on the other hand, has been used to fabricate a two-layer diffractive processor with a lateral feature size of 3 μm and an axial step size of ~125 nm. In contrast, our method demonstrated a lateral feature size of 714 nm, an axial step size of ~100 nm, and 16 phase levels per diffractive feature. Moreover, it supports wafer-level multi-layer fabrication, accommodating ~0.5 billion nanoscale phase features per wafer. This capability ensures enhanced repeatability, precise fabrication consistency, and scalability for mass production, setting a new benchmark for visible diffractive optical processors. Therefore, these combined advantages support more complex and precise designs than previously possible.

**EXPERIMENTAL DEMONSTRATION OF BROADBAND UNIDIRECTIONAL VISIBLE IMAGER**

To experimentally validate our fabricated broadband unidirectional imager, we implemented a setup comprising a tunable laser source, an array of test objects (never used in training), the fabricated multi-layer diffractive design, and a color CMOS image sensor, as illustrated in **Fig. 5a**. We utilized a supercontinuum light source to achieve multispectral illumination, with each spectral band filtered down to ~5 nm bandwidth by an acousto-optic tunable filter. In our experiments, we utilized three narrow spectral lines (~467.5 nm, ~525 nm, and ~627.5 nm) to create multi-color illumination. For the test objects, 100 binary transmittance patterns from the MNIST dataset were fabricated using EBL (see the Methods section). For precise alignment and multi-object imaging, we employed a 3D positioning stage to adjust the x-y-z position of the sample. A 3D-printed holder maintained a 500 μm separation between the diffractive layer and the sensor, and a rotational stage ensured accurate vertical alignment of the laser beam to the sensor. **Figure 5b** shows a photograph of the assembled experimental setup.

As shown in **Fig. 6a**, the experimental measurements aligned well with our simulation results. The input patterns in the forward direction were successfully reproduced across the three illumination wavelengths, while the input patterns in the backward direction were significantly suppressed and distorted, as desired. Quantitative performance evaluations using the PCC metric, shown in **Fig. 6b-c**, yielded a forward average PCC of $0.889 \pm 0.036$ and a backward average PCC of $0.546 \pm 0.041$ in simulations, compared to a forward average PCC of $0.694 \pm 0.052$ and a backward average PCC of $0.435 \pm 0.094$ in our experiments across the three tested wavelengths. These experimental results further confirm the successful design of the broadband unidirectional imager within the visible spectrum.

**CONCLUSIONS**

In this work, we employed unidirectional imaging as a testbed to illustrate the visible design and functionality of wafer-scale diffractive optical processors with approximately 0.5 billion phase features fabricated on the same 6-inch wafer with 16 levels of phase per feature. Importantly, the demonstrated system can be seamlessly adapted to a variety of applications, including unidirectional image magnification[56], classification[42], and operations under incoherent or partially coherent illumination[57,58].

A critical factor for realizing a practical diffractive optical processor is controlling the misalignments between the input/output FOVs and the diffractive layers in a 3D topology. Using specialized alignment marks in our fabrication process, we reduced lateral misalignments between the top and bottom surfaces to < 3 μm (see the Methods section). Although this residual misalignment is small, it could still affect the phase modulation accuracy in our broadband unidirectional imager. The performance degradation from such misalignments is mitigated by applying a vaccination strategy[18,25,31]. To "vaccinate" a diffractive optical processor, random 3D misalignments are modeled in the optical forward model during the training stage so that the diffractive processor learns to adapt to these imperfections. This approach has been proven to significantly improve the robustness of diffractive optical processors against misalignments and fabrication imperfections, and the same strategy was applied to the experimental demonstration of broadband unidirectional imagers reported in **Figs. 1** and **6** (see the Methods section for details).

One of the key advantages of our fabrication method is that it allows our diffractive processor designs to integrate seamlessly with electronic components by utilizing silicon photonics[59,60] to minimize alignment and thermal management challenges. Such an on-chip integrated approach not only enhances performance but also reduces the complexity and footprint of the hybrid system.

In conclusion, we developed a broadband diffractive unidirectional imaging framework that works in the visible spectrum, utilizing HPFS for its exceptional transparency and thermal stability. This system demonstrates robustness across a broad wavelength range and maintains alignment tolerance. Our wafer-level nano-fabrication technique for the creation of diffractive optical processors enables scalability to large FOV applications and multi-tasking; it also allows for seamless integration with electronic components such as CMOS image sensors. This diffractive imaging framework holds significant promise to advance intelligent imaging and sensing applications in the visible spectrum using passive optical elements.

**METHODS**

**Optical forward model of a broadband unidirectional visible imager**

Our diffractive unidirectional imager design consists of $K$ consecutive diffractive layers, each containing thousands of precisely positioned diffractive features. In the numerical forward model, these layers are treated as thin planar structures that modulate incident coherent light with complex transmission functions. For any given $s^{\text{th}}$ diffractive feature on the $l^{\text{th}}$ layer located at $(x_s, y_s, z_l)$, its complex-valued transmission coefficient, dependent on the material thickness value $h_s^l$, can be expressed by the following equation:

$$t(x_s, y_s, z_l; \lambda) = \exp\left(\frac{-2\pi\kappa(\lambda)h_s^l}{\lambda}\right) \exp\left(\frac{-j2\pi(n(\lambda) - n_{\text{air}})h_s^l}{\lambda}\right) \quad (1).$$

Here, $n(\lambda)$ and $\kappa(\lambda)$ are the real and imaginary components of the material's complex refractive index $\tilde{n}(\lambda)$, i.e., $\tilde{n}(\lambda) = n(\lambda) + j\kappa(\lambda)$. For all the diffractive unidirectional imaging designs reported in this paper, we selected HPFS as the material of the diffractive layers[61], with the refractive index curve $n(\lambda)$ provided in **Supplementary Fig. 3**. As HPFS exhibits negligible absorption in the visible range, $\kappa(\lambda)$ was assumed to be 0. The thickness value $h$ of each diffractive element is composed of two parts: a learnable thickness $h_{\text{learnable}}$ and a base thickness $h_{\text{base}}$, such that:

$$h = h_{\text{learnable}} + h_{\text{base}} \quad (2).$$

Here, $h_{learnable}$ is the tunable thickness value of each diffractive feature optimized during the training process and is constrained within the range [0, $h_{max}$]; $h_{base}$ is a fixed value representing the base thickness that acts as the substrate support for the diffractive features, which was empirically chosen as 200 nm. In this paper, $h_{max}$ was set as 1649.5 nm, corresponding to a full phase modulation range from 0 to $2\pi$ for the longest wavelength of interest ($\lambda_1$). Here, the trainable thickness quantization is set to 16 levels based on the fabrication constraint.

To numerically model free-space propagation of coherent light between the diffractive layers, we employed the Rayleigh-Sommerfeld scalar diffraction theory. Each $s^{th}$ diffractive feature on the $l^{th}$ layer at $(x_s, y_s, z_l)$ is defined as the source of a secondary wave, generating a complex field at wavelength $\lambda$ given by the equation:

$$w_s^l(x, y, z; \lambda) = \frac{z - z_l}{(r_s^l)^2}\left(\frac{1}{2\pi r_s^l} + \frac{n}{j\lambda}\right)\exp\left(\frac{j2\pi n r_s^l}{\lambda}\right) \quad (3),$$

where $r_s^l = \sqrt{(x - x_s)^2 + (y - y_s)^2 + (z - z_l)^2}$. These secondary waves propagate to the next layer (the $(l+1)^{th}$ layer), and the optical field that reaches the $p^{th}$ diffractive feature in the $(l+1)^{th}$ layer, located at $(x_p, y_p, z_{l+1})$, can be computed by the convolution of the complex amplitude $u_s^l$ from the previous layer with the impulse response function $w_s^l(x_p, y_p, z_{l+1}; \lambda)$. The resulting field is then modulated by the transmission function $t(x_p, y_p, z_{l+1}; \lambda)$ of the $(l+1)^{th}$ diffractive layer, which can be expressed as:

$$u_p^{l+1}(x, y, z; \lambda) = t(x_p, y_p, z_p; \lambda)\sum_s u_s^l w_s^l(x_p, y_p, z_{l+1}; \lambda) \quad (4).$$

For all the diffractive unidirectional visible imaging designs reported in this paper, the sizes of the input FOV and the output FOV were both set to be 360 × 360 μm. The input/output FOV consists of $N_x \times N_y$ = 28 × 28 pixels, resulting in each output pixel having a size of 12.85 μm × 12.85 μm. To achieve the unidirectional imaging task, we designed the diffractive imager to possess 512 × 512 diffractive features per layer. As the diffractive feature has a size of 714 nm, each layer has a total size of ~366 × 366 μm. The axial distance between the diffractive layers was set to 1,000 μm, while the axial distances from the input plane to the first diffractive layer and from the last diffractive layer to the output plane were both set to 500 μm. In the numerical simulations of all the diffractive unidirectional imaging designs, the spatial sampling period of the simulated complex fields was set to 238 nm. Additionally, during training, we introduced random axial shifts within a range of [-20, 20] μm and random lateral shifts within a range of [-3, 3] μm between the layers. These perturbations were implemented to mitigate the impact of experimental fabrication imperfections and misalignments.

**Training data preparation and other implementation details**

To optimize the diffractive models presented in this study, we utilized a training dataset consisting of 55,000 images from the MNIST handwritten digits. An image augmentation strategy was employed during training to improve the models' generalization capabilities. This strategy included random translation and flipping operations (up-down and left-right) applied to the input images, implemented using the RandomAffine function in PyTorch. The translation range was uniformly sampled within [−5, 5] pixels. Additionally, each flipping operation was performed with a probability of 0.5.

All the diffractive unidirectional imager models used in this work were trained using PyTorch (version 2.5.0, Meta Platform Inc.). We selected the AdamW optimizer[64], and its parameters were taken as the default values and kept identical in each model. The batch size was set as 8. The learning rate, starting from an initial value of 0.03, was set to decay at a rate of 0.5 every 10 epochs, respectively. The training of the diffractive models was performed with 20 epochs. For the training of our diffractive models, we used a workstation with a GeForce GTX 3090Ti graphical processing unit (Nvidia Inc.) and Core i9-11900 central processing unit (Intel Inc.) and 128 GB of RAM, running on Windows 10 operating system (Microsoft Inc.). The typical time required for training a diffractive broadband unidirectional visible imager is ~32 hours.

**Nano-fabrication details**

The unidirectional visible imager design shown in **Fig. 1b** is fabricated on a 6-inch HPFS substrate. This device features a dual-sided diffractive design, with two diffractive layers patterned on opposite surfaces of the substrate. Each diffractive feature in these layers has 16 discrete phase levels, achieving a 4-bit depth phase to cover the 0 to $2\pi$ range at all target illumination wavelengths in the visible spectrum.

To ensure precise alignment across multiple processing steps, specialized micro- and nanoscale alignment marks were patterned on the wafer surfaces. These marks serve as reference points for lithography tools, enabling highly accurate alignment of sequential layers. Even when the wafer is moved, processed, and reloaded, these alignment marks ensure maximum misalignment tolerance is minimized to 3 μm between the top and bottom devices, thus achieving the high precision required for the design.

HPFS was selected as the core material for fabricating our diffractive unidirectional imager due to its extraordinarily low coefficient of thermal expansion (CTE), as shown in **Extended Data Table 2**. This property ensures that even under significant temperature changes, the dimensions and structural integrity of the diffractive features remain stable, preserving the desired optical output without deformation, expansion, or shrinkage.

The input objects shown in **Fig. 6a** were fabricated on a glass slide using EBL and a metal lift-off process. First, the e-beam resist polymethyl methacrylate (PMMA) was spin-coated onto a glass slide and baked on a hot plate at 180 °C for 5 minutes. After e-beam exposure, the resist was developed in a 3:1 solution of methyl isobutyl ketone (MIBK) and isopropyl alcohol (IPA) for 90 seconds, followed by a rinse in IPA for 30 seconds. Subsequently, a 100 nm layer of aluminum was deposited on the sample via magnetron sputtering. The resist was then removed using N-Methyl-2-pyrrolidone (NMP) in an ultrasonic bath at 80 °C. Finally, after the lift-off process, the sample was rinsed for surface cleaning with IPA. This method allowed us to create hundreds of binary amplitude input objects on a single slide.

**Experimental setup**

The experimental setup shown in **Fig. 5** incorporated a supercontinuum laser light source (WhiteLase-Micro; Fianium Ltd, Southampton, UK) to provide multispectral illumination, a microscope slide holder (MAX3SLH; Thorlabs, Inc.) equipped with a sample clip to securely hold the input object slide, and a 3D positioning stage (MAX606; Thorlabs, Inc.) to enable alignment of the object slide. A color CMOS image sensor chip (16.4 MP resolution, 1.12 μm pixel size, Sony Corp., Japan) was employed to capture the image patterns, and a PC controlled the operations of the entire setup. The 3D positioning stage provided precise control over the 500 μm separation between the sample slide and the first diffractive layer, as well as lateral adjustments to position different input objects on the slide. A custom-designed holder, fabricated with a 3D printer (Objet30 Pro, Stratasys), stabilized the diffractive design and maintained a 500 μm separation between the second diffractive layer and the sensor.

In the experimental demonstration of unidirectional visible imaging, we simultaneously illuminated the sample with three distinct wavelengths to achieve multispectral imaging. To address the spectral crosstalk errors among the color channels in the RGB image sensor, we applied a demosaicing algorithm[65,66]. This crosstalk correction is computed by the following equation:

$$\begin{bmatrix} U_R \\ U_G \\ U_B \end{bmatrix} = \mathbf{W} \times \begin{bmatrix} U_{R\_ori} \\ U_{G_1\_ori} \\ U_{G_2\_ori} \\ U_{B\_ori} \end{bmatrix}$$

where $U_{R\_ori}$, $U_{G_1\_ori}$, $U_{G_2\_ori}$ and $U_{B\_ori}$ denote the original patterns captured by the image sensor, $\mathbf{W}$ represents a 3 × 4 crosstalk matrix obtained through experimental calibration specific to the RGB sensor chip, and $U_R$, $U_G$, and $U_B$ are the corrected (R, G, B) image patterns.

**Supplementary Information:** This file contains:

- Supplementary Figures 1-3.
- Extended Data Tables 1-2
- Training loss functions and performance evaluation metrics

**FIGURES**

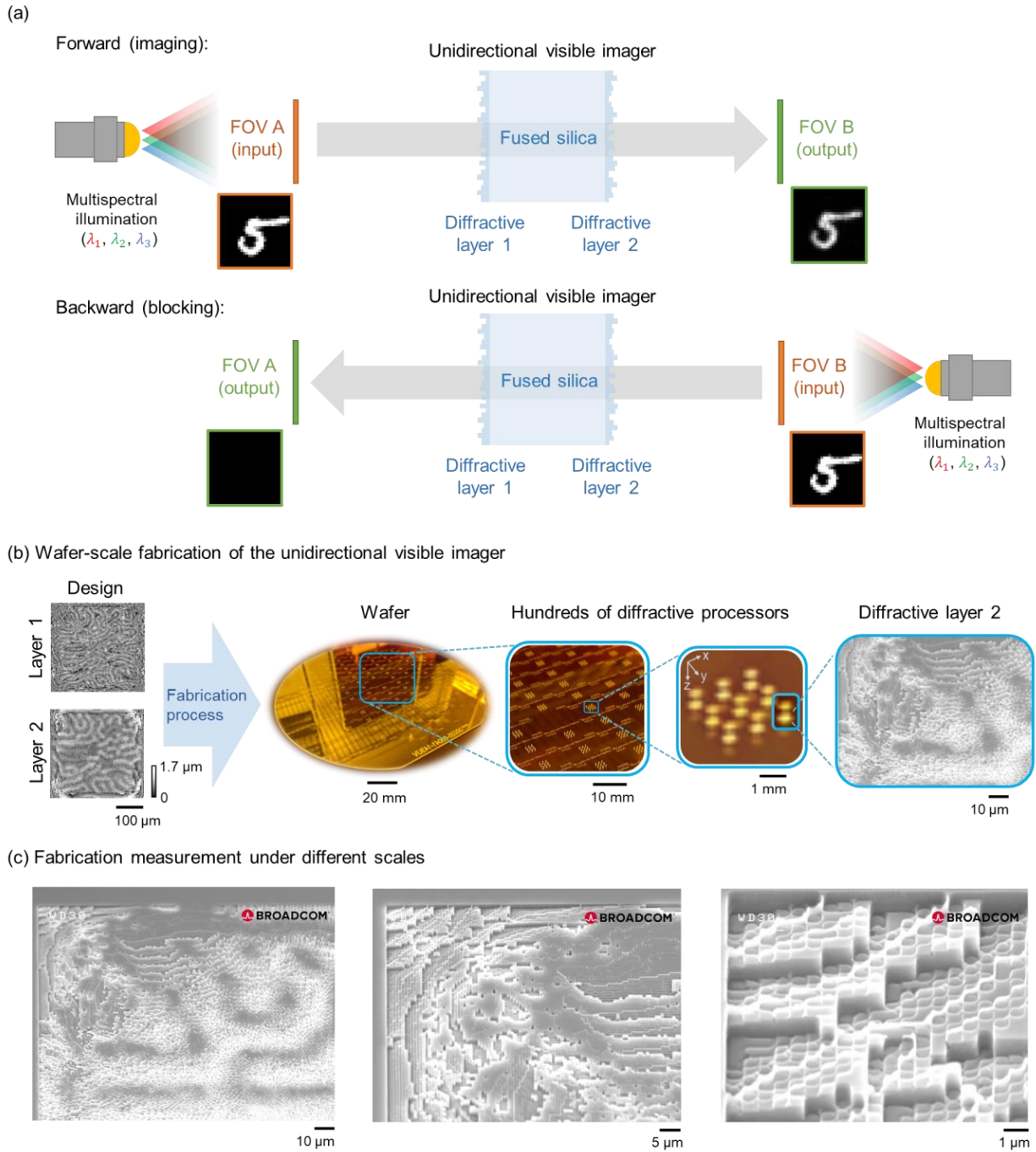

**Fig. 1 Illustration of broadband unidirectional visible imagers using wafer-scale fabrication of multi-layer diffractive optical processors.** (a) A unidirectional imager reproduces the image in the forward direction (FOV A to FOV B) while blocking the image transmission in the backward direction (FOV B to FOV A). (b) Thickness profiles of the optimized diffractive layers and wafer-scale fabrication of 918 multi-layer diffractive designs on the same 6-inch wafer. (c) SEM images of the diffractive layers at different magnification factors.

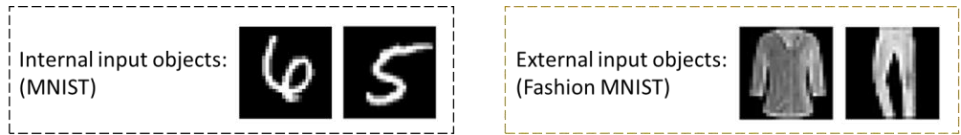

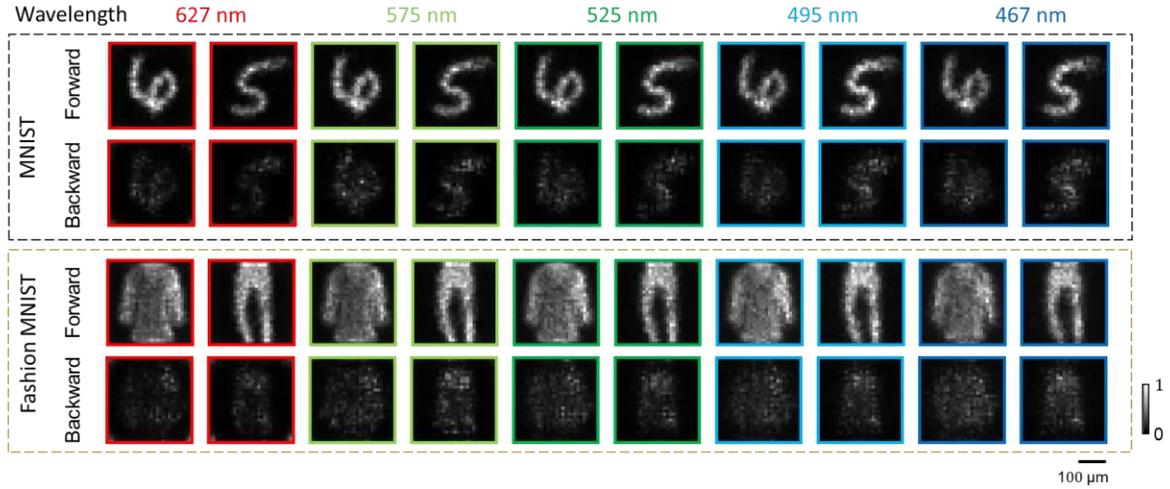

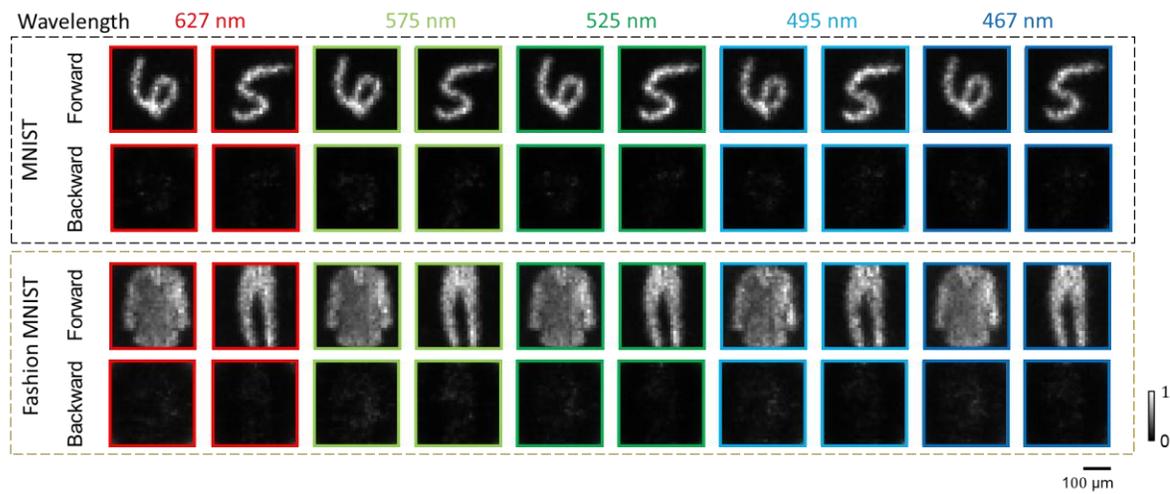

**Fig. 2 Blind testing results of broadband unidirectional visible imagers at different illumination wavelengths.** (a) Diffractive output images of both the forward direction and the backward direction using the two-layer unidirectional imager design shown in **Fig. 1b**. (b) Diffractive output images of both the forward direction and the backward direction using the three-layer unidirectional imager design shown in **Supplementary Fig. 1**.

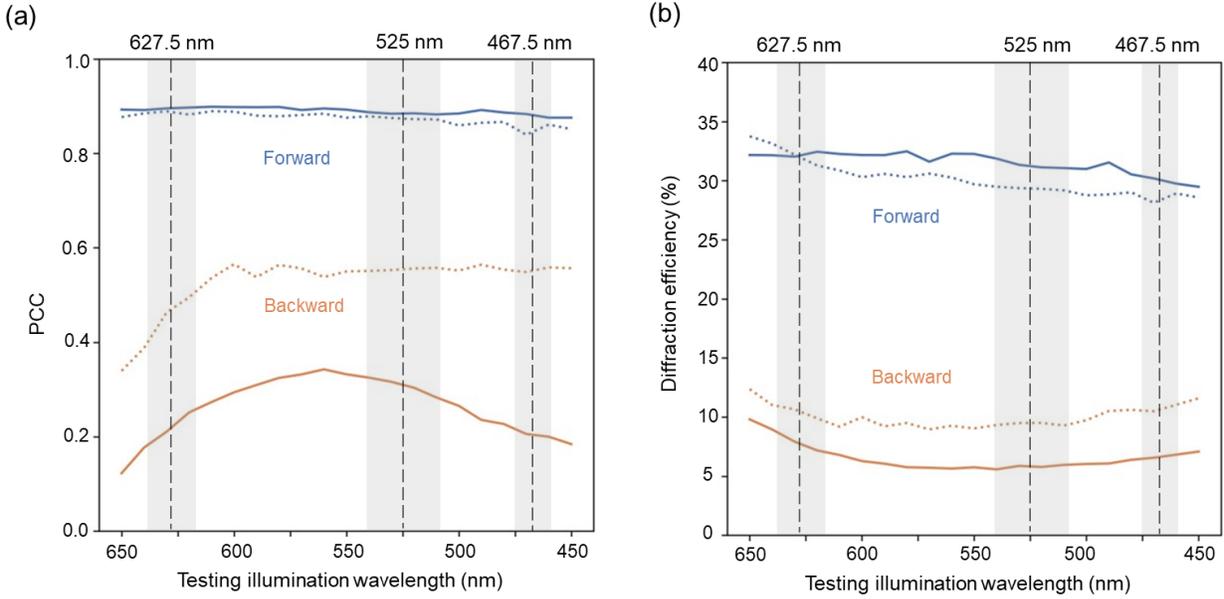

**Fig. 3 Spectral response of broadband unidirectional visible imagers.** (a) Output image PCC values as a function of the illumination wavelength for both the forward and backward directions. (b) Diffraction efficiency as a function of the illumination wavelength for both the forward and backward directions. The dashed curves represent the performance of the two-layer unidirectional imager design shown in **Fig. 1b**, while the solid curves represent the performance of the three-layer unidirectional imager design shown in **Supplementary Fig. 1**. The gray areas mark the illumination wavelengths used during the training of the diffractive optical processors.

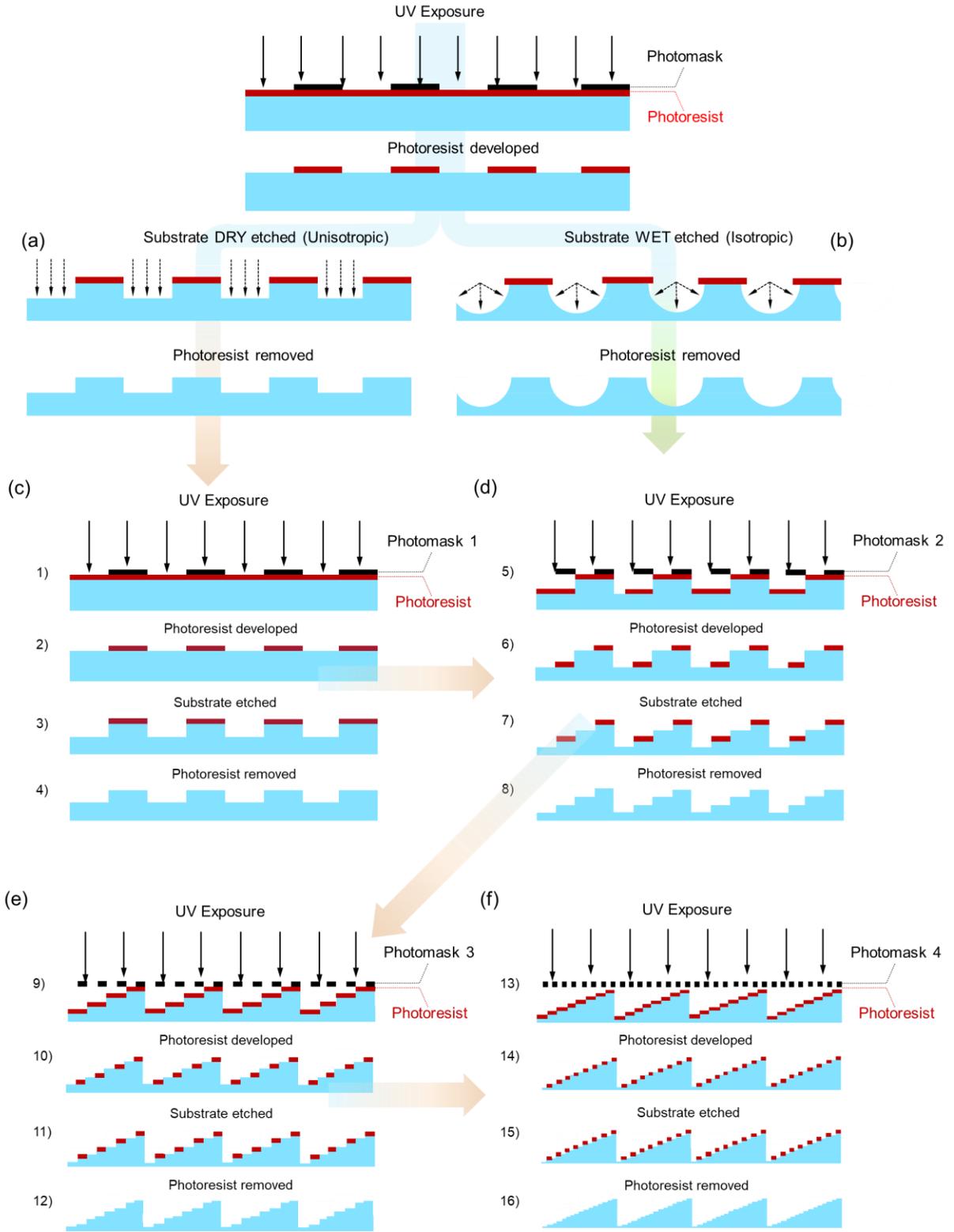

**Fig. 4 Lithography process for the fabrication of diffractive optical layers.** (a) Illustration of dry etching versus (b) wet etching. (c) Fabrication sequence to obtain a 2-level diffractive surface. (d) Subsequent fabrication to generate a 4-level diffractive surface. (e) Subsequent fabrication to generate an 8-level diffractive surface. (f) Subsequent fabrication to generate a 16-level diffractive surface. (c)-(f) depict a conceptual fabrication of a 16-level diffractive surface designed using the binary optics model (2N).

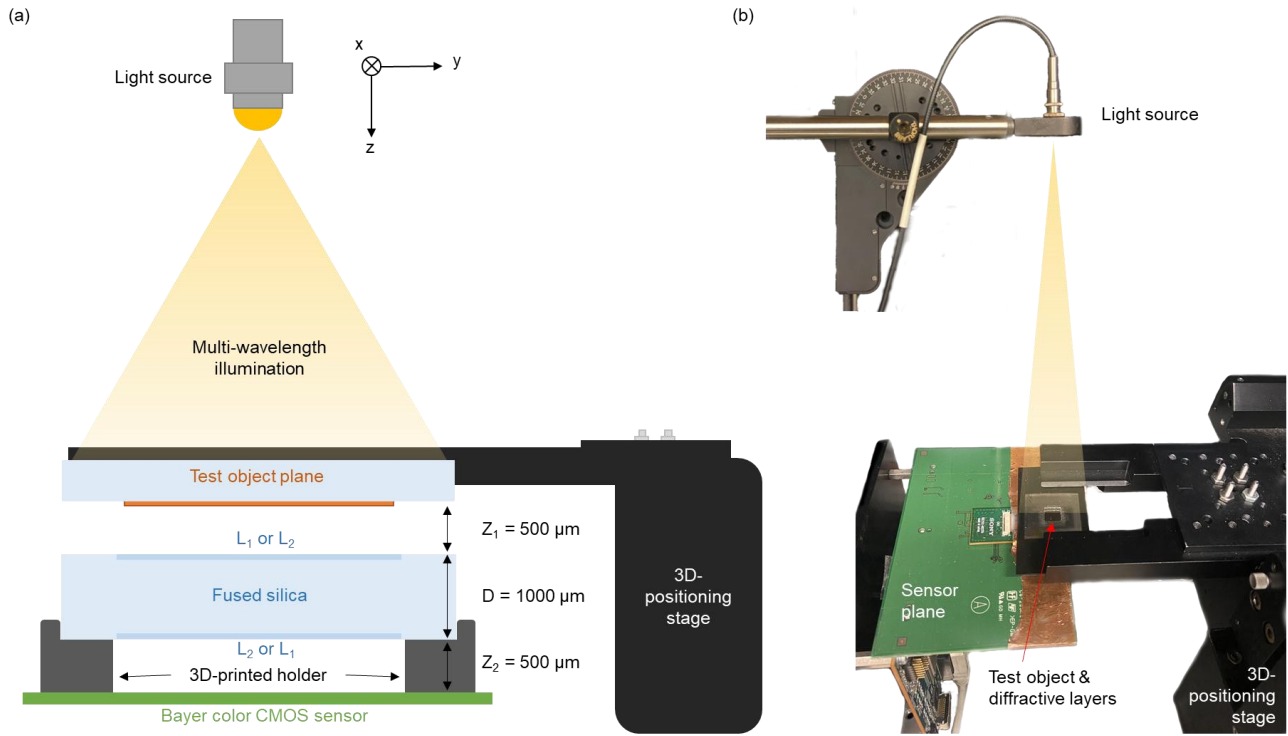

**Fig. 5 Experimental setup of broadband unidirectional visible imaging.** (a) Illustration of the experimental setup, including the tunable laser source, the test object, the fabricated diffractive layers and the image sensor. (b) A photo of the experimental setup.

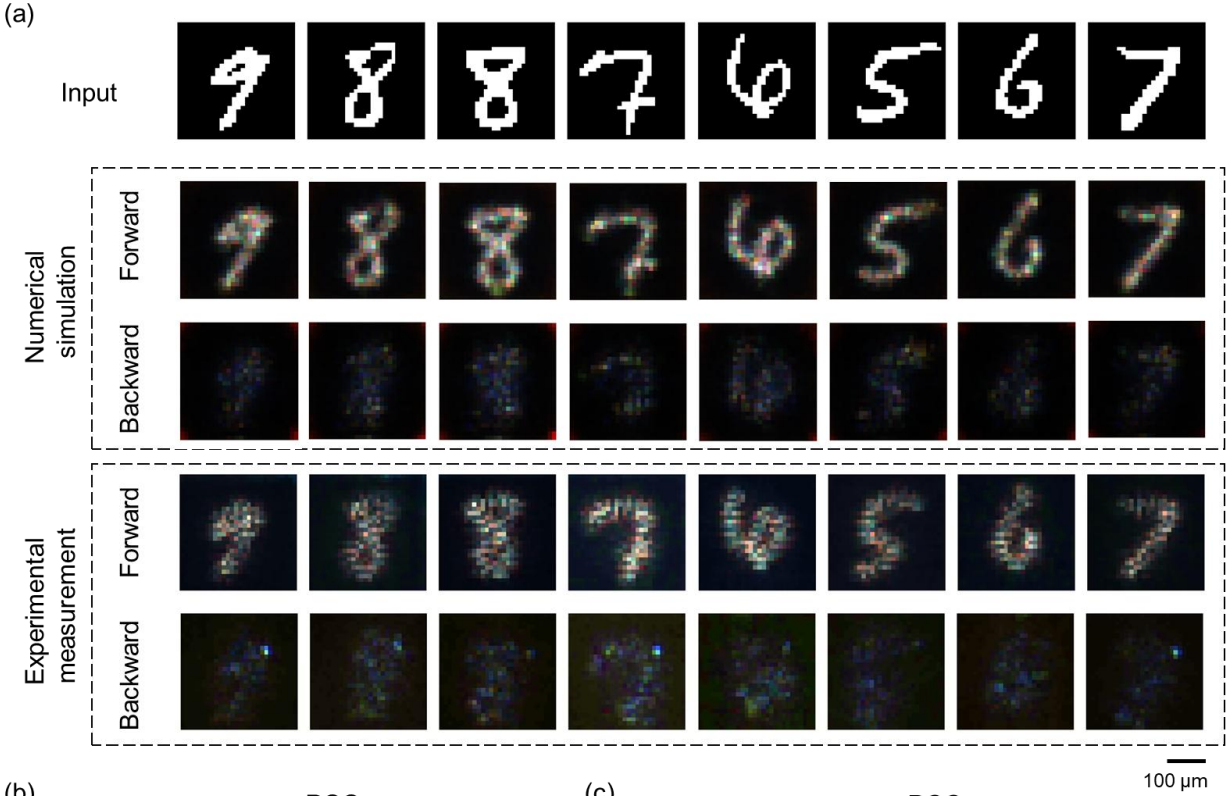

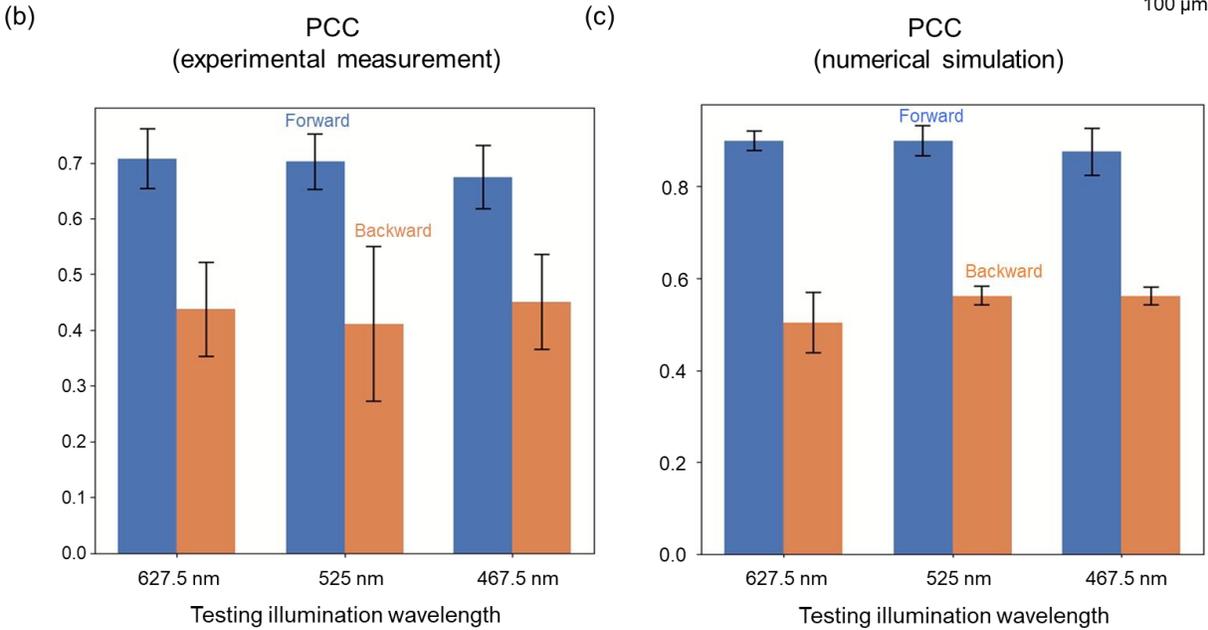

**Fig. 6 Experimental results of broadband unidirectional visible imaging.** (a) Experimentally measured diffractive outputs of both the forward and the backward directions, along with the numerically simulated outputs. (b) Output image PCC values of the forward and the backward directions from the experimental measurements at three illumination wavelengths. (c) Output image PCC values of the forward and the backward directions from the numerical simulations at three illumination wavelengths. These values are calculated across test output images, reported with the mean values and the standard deviations shown as error bars.